\documentclass[journal,onecolumn]{IEEEtran}
\ifCLASSINFOpdf
\else
\fi
\usepackage{graphicx}%
\usepackage{multirow}%
\usepackage{amsmath,amssymb,amsfonts}%
\usepackage[numbers]{natbib}
\usepackage{amsthm}%
\usepackage{mathrsfs}%
\usepackage{xcolor}%
\usepackage{textcomp}%
\usepackage{manyfoot}%
\usepackage{makecell}
\usepackage{booktabs}%
\usepackage{algorithm}%
\usepackage{algorithmicx}%
\usepackage{algpseudocode}%
\usepackage{listings}%
\usepackage{caption}
\usepackage[table,xcdraw]{xcolor}
\usepackage{framed}
\usepackage{enumitem}
\usepackage{soul}
\usepackage{mdframed}
\usepackage{hyperref}

\usepackage{fixltx2e}
\definecolor{formalshade}{rgb}{0.95,0.95,1}
\definecolor{darkblue}{rgb}{0.145, 0.118, 0.580}

\newenvironment{formal}{%
  \MakeFramed{\advance\hsize-\width\FrameRestore}%
  \noindent\hspace{-4.55pt}

  \vspace{2pt}\vspace{2pt}%
}
{%
  \vspace{2pt}\endMakeFramed%
}

\begin{document}
%
\title{AI/ML Based Detection and Categorization of Covert Communication in IPv6 Network}

%
%
%

\author{
Mohammad Wali Ur Rahman\IEEEauthorrefmark{1},
Yu-Zheng Lin\IEEEauthorrefmark{1},
Carter Weeks\IEEEauthorrefmark{2},
David Ruddell\IEEEauthorrefmark{3},
Jeff Gabriellini\IEEEauthorrefmark{3},
Bill Hayes\IEEEauthorrefmark{2},
Salim Hariri\IEEEauthorrefmark{1}, Pratik Satam\IEEEauthorrefmark{1}\IEEEauthorrefmark{3},
Edward V. Ziegler Jr.\IEEEauthorrefmark{4}\\
\IEEEauthorblockA{\IEEEauthorrefmark{1}Department of Electrical and Computer Engineering, University of Arizona, Tucson, AZ 85721, USA}\\
\IEEEauthorblockA{\IEEEauthorrefmark{2}Department of Management Information Systems, University of Arizona, Tucson, AZ 85721, USA}\\
\IEEEauthorblockA{\IEEEauthorrefmark{3}Department of Systems and Industrial Engineering, University of Arizona, Tucson, AZ 85721, USA}\\
\IEEEauthorblockA{\IEEEauthorrefmark{4}Laboratory for Advanced Cybersecurity Research, National Security Agency, USA}
}

%
%

\markboth{Accepted by Springer Cybersecurity (Sep. 14 2025)}%
{Shell \MakeLowercase{\textit{et al.}}: Bare Demo of IEEEtran.cls for IEEE Journals}
%



\maketitle

\begin{abstract}
The flexibility and complexity of IPv6 extension headers allow attackers to create covert channels or bypass security mechanisms, leading to potential data breaches or system compromises. The mature development of machine learning has become the primary detection technology option used to mitigate covert communication threats. However, the complexity of detecting covert communication, evolving injection techniques, and scarcity of data make building machine-learning models challenging. In previous related research, machine learning has shown good performance in detecting covert communications, but oversimplified attack scenario assumptions cannot represent the complexity of modern covert technologies and make it easier for machine learning models to detect covert communications. To bridge this gap, in this study, we analyzed the packet structure and network traffic behavior of IPv6, used encryption algorithms, and performed covert communication injection without changing network packet behavior to get closer to real attack scenarios. In addition to analyzing and injecting methods for covert communications, this study also uses comprehensive machine learning techniques to train the model proposed in this study to detect threats, including traditional decision trees such as random forests and gradient boosting, as well as complex neural network architectures such as CNNs and LSTMs, to achieve detection accuracy of over 90\%. This study details the methods used for dataset augmentation and the comparative performance of the applied models, reinforcing insights into the adaptability and resilience of the machine learning application in IPv6 covert communication. We further introduce a Generative AI-driven script refinement framework, leveraging prompt engineering as a preliminary exploration of how generative agents can assist in covert communication detection and model enhancement.
\end{abstract}


%
\IEEEpeerreviewmaketitle

\section{Introduction}
\label{sec:introduction}
Internet Protocol version 6 (IPv6) is the latest developed version of the Internet Protocol with a larger addressing space. It was launched to solve the problem of IPv4 addresses being exhausted. Several improvements are designed to improve network security and efficiency and apply to various devices. However, extension headers in IPv6 allow attackers to inject covert traffic, which increases the risk of data leakage or injects malware that is difficult to detect. Therefore, exploring methods for detecting IPv6 covert communications is crucial in emerging network environments.

In recent years, advances in computing power and significant research investment have propelled machine learning to the forefront of cybersecurity, particularly in the classification and analysis of network traffic \mbox{\cite{pacheco2018towards}}. As a result, machine learning techniques have become the preferred approach for detecting packets containing covert communications and pinpointing the specific header fields used for such injections in IPv6 traffic. The primary objective is to mitigate the risks posed by covert communication attacks within IPv6-enabled networks. However, with the rapid evolution of attack techniques, effectively training machine learning models for this purpose demands access to large volumes of high-quality data, as well as a comprehensive understanding of the diverse methods by which covert communications can be embedded within IPv6, in order to realistically capture and address emerging threats.

In this study, we adopt the prisoner’s problem framework \cite{simmons1984prisoners} to model a covert communication scenario, where Alice attempts to transmit a hidden message to Bob while an adversary, Wendy, monitors the traffic for evidence of covert activity. To support machine learning-based detection, we construct a realistic IPv6 dataset with embedded covert communications. The dataset is designed to reflect practical usage patterns and structural variability in IPv6 header fields, thereby capturing the complexity and subtlety of real-world covert channels.

Using this dataset, we evaluate a range of machine learning models, including random forests, gradient boosting, convolutional neural networks (CNN), and long short-term memory networks (LSTM). These models are assessed for their ability to detect covert communications embedded within IPv6 packets. The experimental results offer insights into the relative effectiveness of different algorithms and inform the selection of models suitable for IPv6 traffic analysis. This work contributes to the development of machine learning-based intrusion detection systems (IDS) \mbox{\cite{pacheco2020artificial,satam2020wids}} by providing both a representative data set and an empirical benchmark for covert channel detection in IPv6 environments.

\section{Related Works}
While the security of IPv6 has been extensively studied in various domains, the specific issue of covert communications within the protocol remains comparatively underexplored. Prior research has primarily concentrated on the feasibility of embedding hidden data in IPv6 header fields, a possibility enabled by the protocol’s expanded header structure and increased field flexibility.

A recent survey by Khadse and Dakhane consolidates 2015–2024 work on covert-channel construction and detection, emphasizing bandwidth–undetectability trade-offs and persistent dataset scarcity \cite{khadse2025review}. The Study by Lucena et al. \cite{lucena2005covert} has been foundational, identifying up to 22 covert channels within IPv6. These channels leverage underutilized fields such as the Flow Label and Payload Length, offering stealthy methods for transmitting hidden information across networks while evading standard detection mechanisms. Another notable development is the pcapStego tool by Zuppelli and Caviglione (2021), designed to inject covert communications directly into IPv6 headers, simulating real-world covert channel utilization for research purposes \cite{Zuppelli2021}. Another recent advance is the open-source IPv6CC suite, which provides penetration testers and researchers with multiple methods to inject and evaluate covert channels in IPv6 networks using various header fields and injection strategies \cite{caviglione2022ipv6cc}. IPv6CC enables assessment of organizational security posture against stegomalware and exfiltration threats, and allows for practical benchmarking of detection systems, including their ability to withstand both naive and sophisticated attacks. Suprun et al. propose a Flow Label centric steganographic model secured with ECDH key exchange and ECDSA signatures, reducing encode/decode latency versus earlier CBC-RC6 variants while retaining a 20-bit-per-packet capacity \cite{suprun2025development}. On the detection side, bccstego introduces a scalable eBPF-based framework that inspects IPv6 header fields in real time to collect statistical indicators and reveal covert channel activity \cite{repetto2021bccstego}. By leveraging efficient binning and a low processing footprint, bccstego can be deployed in production environments for proactive detection without adversely impacting legitimate traffic or system resources.

\begin{figure*}[b!]
\centering
\includegraphics[width = \textwidth]{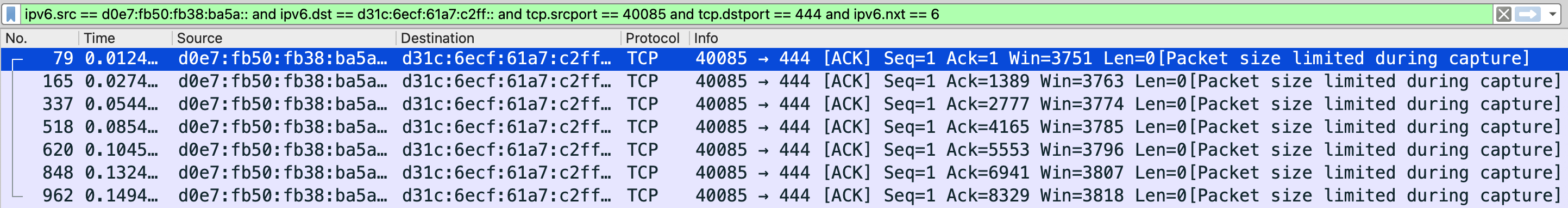}
\caption{Example of covert communication injection by pcapStego}
\label{fig:pcapstego}
\end{figure*}

Despite these advancements, Dua et al.~\cite{Dua2022} reveal a gap in practical applicability, as their study is based on idealized assumptions that limit its relevance to real-world deployment scenarios. Their research employed pcapStego to craft IPv6 packets with embedded covert channels. While their work on a two-layer detection framework combining Deep Neural Networks (DNN) and Support Vector Machines (SVM) represents a significant technical achievement, whether the crafted covert communications packages reflect real-world threats has been questioned. The resulting covert communications are oversimplified and unrealistic, so these features make the covert packet generated by pcapStego easy to detect. For instance, Figure \mbox{\ref{fig:pcapstego}} illustrates a covert communication scenario in which the word “Hello” is embedded within the IPv6 Traffic Class field using the pcapStego tool. Although Wireshark displays these packets as part of a continuous TCP flow, the sequence number has been manually reset to 1. This inconsistency renders the packets invalid from the perspective of the TCP protocol stack and would likely result in connection errors if transmitted over a real network. While injecting the covert channel into the IPv6 header, the pcapStego application created fake traffic that mimicked TCP traffic. As discussed in \cite{houmansadr2013}, this type of traffic is relatively easy to identify. The resulting artifacts, which are unrelated to the content of the injected data, can inadvertently serve as distinguishing features for machine learning algorithms. This reliance on such superficial characteristics undermines the practical relevance of the findings and may lead to an overestimation of detection performance in real-world scenarios.

This criticism is echoed across the literature, where the lack of high-quality, realistic datasets for IPv6 covert communication is a recurring theme. The gap in data realism significantly hampers the development and testing of effective detection mechanisms. Innovative research like Dua et al. (2022) is crucial but must be underpinned by scenarios that accurately reflect the sophistication and subtlety of potential real-world attacks.

Recent field measurements by Mazurcyk et al. demonstrate that multiple covert techniques are already deployed in operational networks and benchmark detection performance across six widely used hiding tools. These real-world observations underscore the need for datasets and detection methods that reflect live traffic conditions \mbox{\cite{mazurczyk2019ipv6}}. Additionally, Dua et al. (2023) introduced the DICCh-D framework, using deep neural networks on blended CAIDA normal traffic and realistic covert samples, to improve detection accuracy to over 99.5\% \mbox{\cite{dua2022dicch}}. Recent work by Bedi et al. \mbox{\cite{bedi2023spyipv6}} introduced SPYIPv6, a two-layer framework that uses KNN-based classification to not only detect covert IPv6 packets but also accurately localize the header fields (or combinations thereof) used for data hiding. SPYIPv6 demonstrates high accuracy (99.85\%) and fast detection on blended CAIDA and synthetically generated covert traffic, representing a significant advancement in practical detection and localization of multi-field IPv6 covert channels.

A complementary rule-based approach, CC-Guard \mbox{\cite{wang2022cc}}, matches IPv6 header values against learned normal-traffic templates and achieves real-time detection of ten storage channels with negligible runtime overhead. Further advancing the field, Zhang et al. \mbox{\cite{zhang2024self}} introduce a self‑attention mechanism‑based model for detecting multi‑field covert channels in IPv6, covering both base and extension headers. Their model employs multi‑head attention to correlate sub‑field embeddings and accurately locate covert data across 23 channel types. Evaluated on a publicly available IPv6 covert channel dataset, it demonstrated precision of 97. 13\% with a false positive rate of just 6. 3\%, significantly outperforming previous approaches such as BNSCNN and DICChD in both detection breadth and localization accuracy. These studies reinforce the importance of quality and representativeness in covert traffic datasets.

Furthermore, the broader field of IPv6 security must contend with evolving threats and the continuous adaptation of attack vectors. As IPv6 adoption increases, so does the complexity of maintaining robust security protocols. Current research must evolve beyond theoretical analyses and synthetic datasets to incorporate more dynamic, realistic testing environments. This will improve the accuracy of detection algorithms and enhance the overall security posture of network systems operating under the IPv6 protocol \cite{caicedo2009ipv6}.

\section{Methodology}
\subsection{Threat Modelling}
\begin{figure}[b!]
\centering
\includegraphics[width = 0.75\columnwidth]{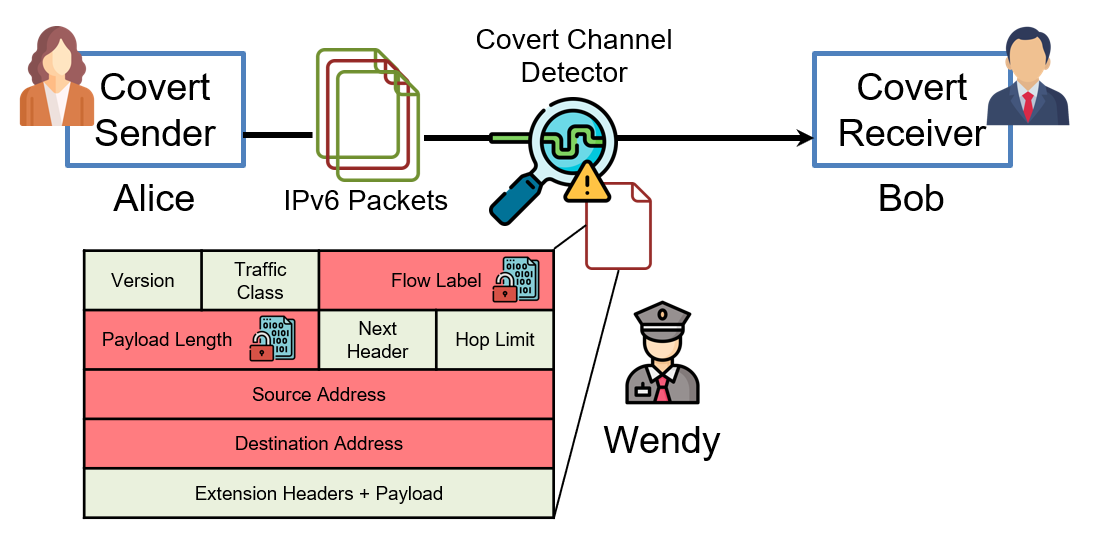}
\caption{The prisoner's scenario with IPv6 covert communication}
\label{fig:prisoner_scenario}
\end{figure}

As shown in Figure \ref{fig:prisoner_scenario}, our threat modeling is motivated by the idea of the Prisoners' Problem and the application of Subliminal Channels \cite{simmons1984prisoners}. In this scenario, two parties, Alice and Bob, want to communicate secretly through IPv6 packets so that Wendy, the warden, will not be able to intercept their communication. Manipulation of several IPv6 header fields, including the Flow Label, Payload Length, and the Source/Destination Address Spaces, may enable covert communication \cite{caviglione2021code,mazurczyk2019ipv6,ackerman2015covert,ipv6security2011}. These fields might contain encrypted data or patterns that a novice observer or a simple detection system would not immediately recognize as anomalous.

Our primary goal is to enable Wendy to recognize these secret communications and, and retrieve the hidden information with machine learning models. These models applies sophisticated machine learning techniques to identify discrepancies that indicate covert activities using patterns found from the analysis of these header fields \cite{elsadig2022covert, caviglione2021trends}.

\begin{figure*}[t]
\includegraphics[width = 1.025\textwidth]{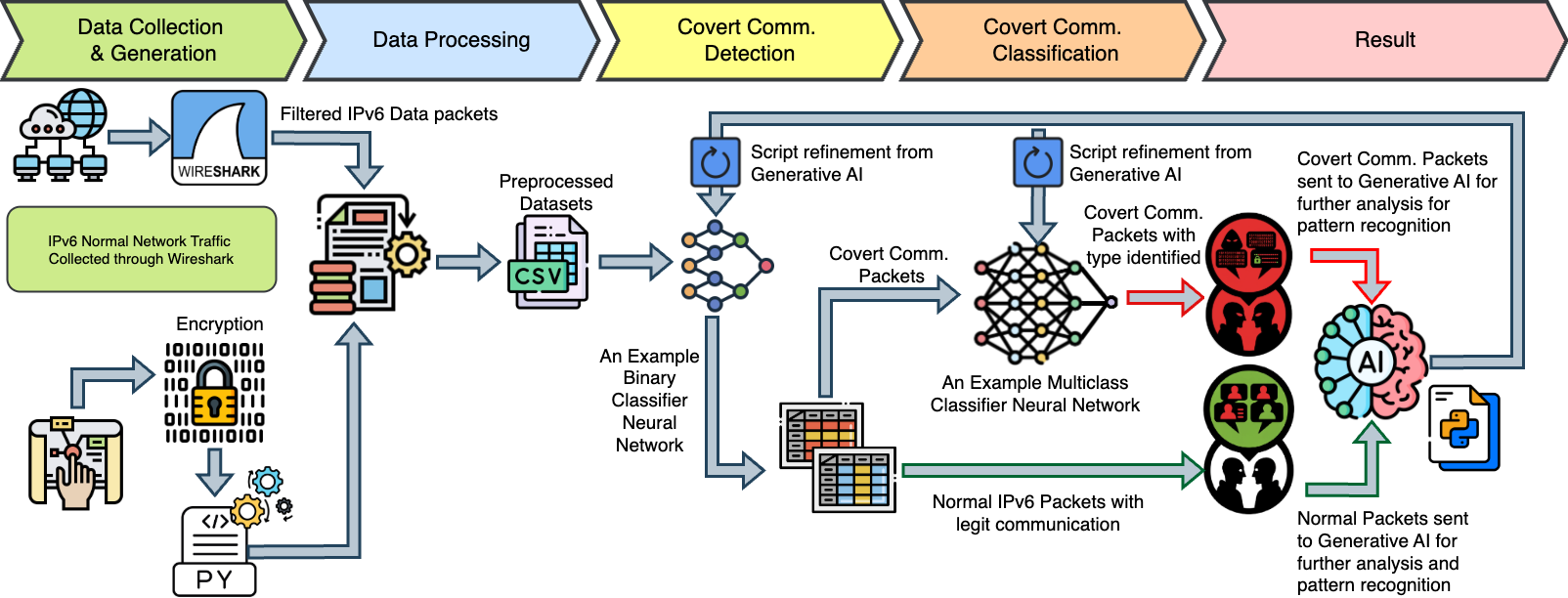}
\caption{IPv6 Covert Communication Detection and Classification Framework}
\label{fig:insure_framework}
\end{figure*}

\subsection{Operational Pipeline}


Our research employs a comprehensive machine learning pipeline to detect and classify covert communications embedded within IPv6 network traffic, as illustrated in Figure~\ref{fig:insure_framework}. This pipeline encompasses stages such as data collection, encryption, preprocessing, and the application of advanced neural network models. Below, we provide an overview of the five key steps involved in this process. Each step will be discussed in greater detail in the subsequent subsections.

\begin{enumerate}
    \item \textbf{Data Collection and Generation:}  Covert Communication Traffic are Generated and carefully crafted with various encryption methods of different IPv6 fields and python script to automate the process. 
    \item \textbf{Data Processing:} The vulnerable IPv6 fields are selected for analysis through Feature Selection. Standard data processing steps like Normalization or Label Encoding may or may not be performed, depending on the nature of the IPv6 field.
    \item \textbf{Covert Communication Detection:} A Binary Classification Model is trained to classify IPv6 Network Packets based on Normal Behavior. After prediction, the normal and suspected covert communication IPv6 packets are separated, where the Suspected Covert Communication Packets will undergo further analysis.
    \item \textbf{Covert Communication Classification:} After separating the covert communication data packets from the normal ones, the covert communication packets are passed to a multiclass classification model. The multi-class classification model further classifies the covert comm packets, based on the IPv6 fields that were exploited for covert communication
    \item \textbf{Result:} After the complete cycle, the covert communication and normal data packets in IPv6 are sent to the Generative AI agent where the covert communication packets are analyzed to see whether there are more patterns that can be used to identify and categorize covert communication packets, and based on the newly found products and classification results, the python scripts for detection and classification both are revised and refined and the iteration goes on.
\end{enumerate}

\subsubsection{Data Collection and Generation} \label{subsec:data_collection}

Data for this study was meticulously collected and generated to ensure a robust dataset for machine learning analysis. Below are details on the acquisition and generation of both normal and covert traffic:


\begin{itemize}
    \item \textbf{Normal Traffic Acquisition:} Data on normal IPv6 traffic was obtained from the 2019 IPv6 Launch Day Anonymized Internet Traces provided by CAIDA \cite{caida2019}. This dataset offers a thorough view of typical network behavior. The dataset was further processed using the well-known network protocol analyzer Wireshark \cite{beale2006wireshark} to filter and prepare the data for our experiments.

    \item \textbf{Covert Traffic Creation:} To simulate the anomalous or covert class, covert communication traffic was created synthetically. The covert traffic was created using Python scripts that applied different encryption techniques to IPv6 fields. This ensured that the traffic looked authentic. This methodology tests and challenges our machine learning models' detection capabilities.
\end{itemize}

\begin{table}[h]
\centering
\captionsetup{width=1\linewidth, justification=centering}
\begin{tabular}{ll}
\hline
\textbf{Category}                        & \textbf{Count}   \\ \hline
\textbf{Normal}                         & \textbf{411,720} \\ 
\textbf{Covert (Total)}                & \textbf{313,738} \\
\quad-- HopLimit Encoding & 116,628          \\
\quad-- Address Space     & 76,575           \\
\quad-- Length Encrypt    & 75,823           \\
\quad-- Flow Label        & 44,712           \\ \hline
\textbf{Total Packets}                 & \textbf{725,458} \\ \hline
\end{tabular}
\caption{Summary of Training Packets in the Dataset}
\label{tab:combined_dataset_summary}
\end{table}

The dataset used in our machine learning research is summarized in Table \ref{tab:combined_dataset_summary}. 

In addition, the covert packets were further categorized according to the fields in IPv6 packets that were encrypted, providing us insights about the IPv6 fields, that are exploited the most during covert communications.

\subsubsection{Data Processing}
\begin{itemize}
\item\textbf{Filtering and Preprocessing:} Before being preprocessed to make it appropriate for machine learning analysis, the gathered data is filtered to remove unnecessary information. This could involve preparing the data into an organized CSV format for additional processing, cleaning the data, and dealing with missing values. 

\item\textbf{Feature Selection:} We employ domain knowledge and data characteristics to perform feature selection, identifying which IPv6 fields are most indicative of covert communications. This step enhances model accuracy and computational efficiency \cite{gong2025feature}.

\item\textbf{Normalization and Encoding:} The data may be normalized or encoded depending on the nature of the IPv6 fields, to prepare it for use in machine learning models. Normalization scales numeric fields to a uniform scale, while encoding transforms categorical fields into numeric formats \cite{preprocessing2020}.
\end{itemize}

\subsubsection{Covert Communication Detection} 
\begin{itemize}
\item\textbf{Binary Classification:} A neural network for binary classification is trained using a labeled dataset that includes both covert and normal traffic. As a first filter in the detection process, the classifier's job is to discern between typical network activity and possible covert communications. 

\item\textbf{Separation of Data:} After classification, streams of traffic are separated into those that are likely to be normal and those that might contain covert communications. This allows for targeted analysis of suspicious traffic. 
\end{itemize}
\subsubsection{Covert Communication Classification}
\begin{itemize}

\item\textbf{Multiclass Classification:} An in-depth examination employing a multiclass classifier neural network is performed on traffic that has been recognized as possibly covert communication. The sophisticated model has been trained not only to identify the type of communication based on manipulated IPv6 fields, but also to confirm the existence of covert communications. 
\end{itemize}

\subsubsection{Result} 
\begin{itemize}
\item\textbf{Identification and Classification of Covert Packets:} The pipeline's last stage involves identifying and classifying covert IPv6 packets, which are identified and categorized according to their type. This process offers important insights about different IPv6 fields that are used for covert communication. 

\item\textbf{Differentiation from Normal Traffic:} Legitimate IPv6 packets are filtered out as normal traffic, allowing for efficient network operation and focused security analysis on detected anomalies.

\item\textbf{Script Refinement with Generative AI:} Based on the classifiers' performance on detecting and classifying covert communication packets, a dedicated Generative AI agent expert in code generation task will review the results and refine or make adjustments to the scripts in an iterative process until the performance exceeds an expected threshold.
\end{itemize}

\section{IPv6 Vulnerability Assessment}



\subsection{Hop Limit}
    


\begin{figure*}[t]
\centering
\includegraphics[width=\textwidth]{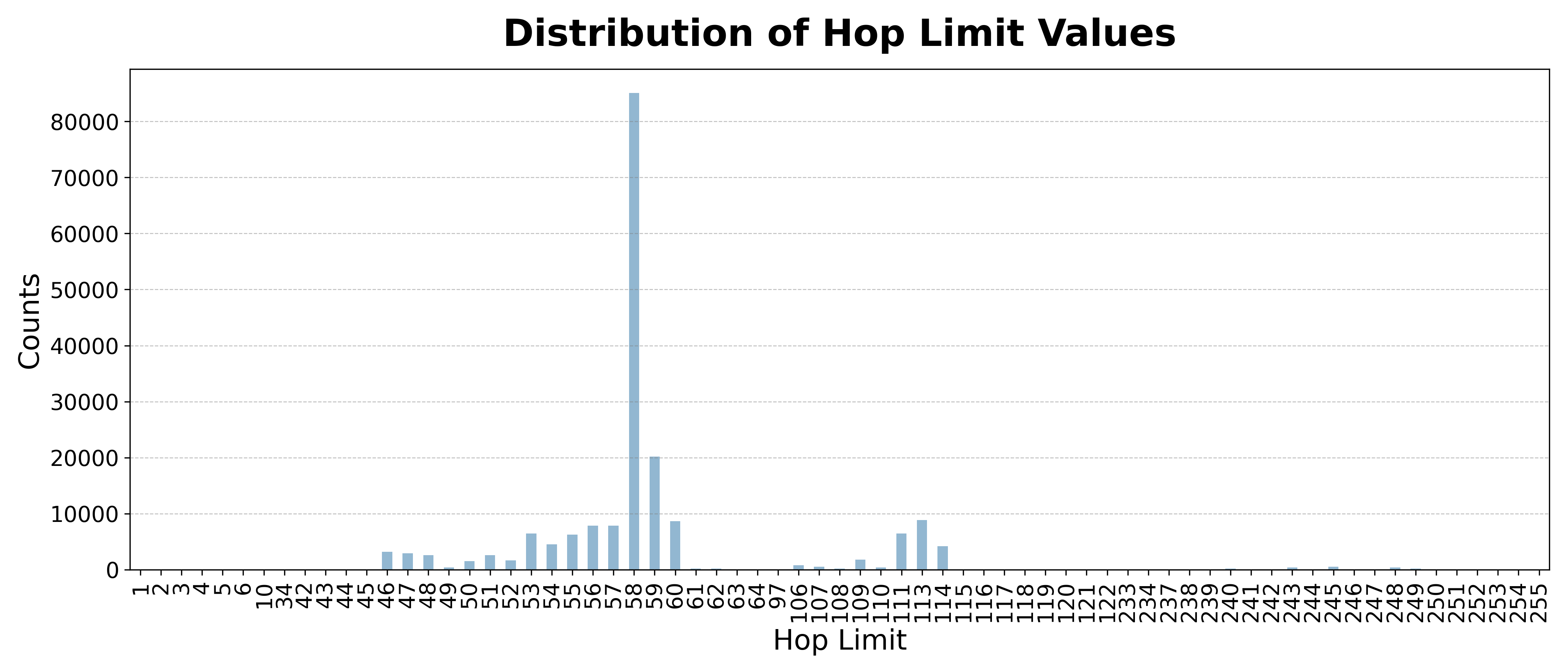}
\caption{Distribution of Hop Limit Values}
\label{fig:hoplimit}
\end{figure*}

Analysis of the distribution of Hop Limit values in the CAIDA dataset reveals a high distribution of counts between 48 and 60 and a lower distribution of counts between 109 and 114. This is indicative of the general use of unix-like operating systems (Linux and BSD) for servers that generate a greater number of packets and the use of Windows operating systems for clients on the Internet, as unix-like operating systems generally default to a Hop Limit of 64 and Windows defaults to a Hop Limit of 128. A smaller set of data between 233 and 255 indicates that some of the traffic was generated by older versions of Linux, which set the Hop Limit to 255 by default, or alterations to the default settings of other operating systems (see Figure \ref{fig:hoplimit}). These characteristics of the Hop Limit place restrictions on how it can be used effectively as a covert channel. As was pointed out in \cite{Dua2022}, the most effective way to use the Hop Limit is to choose between values of 64 and 128 to represent either one or zero to transmit a bit of information with each packet. This is effective as the average number of hops on the internet as found in \cite{Zhang2023} was 9.497, so the decrementing of the value as it traverses the network is unlikely to corrupt the value, as values greater than 64 could be considered a zero and other values a one. It is also possible to encode additional information if one also includes 255 as a possible value, allowing one to encode a ternary bit per packet. 

\subsection{Traffic Class}

Traffic differentiation and network congestion notification are crucial functions of the Traffic Class field in IPv6, comprising the Differentiated Services Code Point (DSCP) and Explicit Congestion Notification (ECN) values. With these elements as our primary focus, our analysis shows that typical traffic patterns vary little. As suggested by \cite{rfc2474} for DSCP and \cite{rfc3168} for ECN standards, the DSCP values primarily exhibit a strong bias towards default settings with slight deviations into other classifications. 

Because the Traffic Class values are so uniform, any change, such as encrypting them, would cause observable irregularities. Using anomaly-based network intrusion detection systems, such departures from accepted norms could be quickly identified. Thus, to preserve the stealth qualities necessary for successful covert communication, we decided not to alter the Traffic Class values in our machine learning study when creating covert communication packets. By making this choice, we can ensure that the generated traffic blends in with typical traffic patterns and that we are simulating realistic network conditions. 

\subsection{Flow Label}

The Flow Label field in IPv6 is designed to provide a means for labeling sequences of packets that belong to the same flow, a concept intended to facilitate packet processing along the path of the flow. Our analysis has highlighted significant patterns in the use of Flow Labels across different protocols which have implications for their potential use in covert communications \cite{rfc6437}.

\subsubsection{Pattern Analysis}
Flow Label values in TCP packets indicate a particular flow and are frequently set to ``0x00000'' or sporadically stay constant. Because of this regularity, any unusual or encrypted Flow Labels within TCP protocols should be simple to identify because they deviate from the norm. Because of this, TCP packets are not the best for concealing secret communications without running the risk of being discovered. 

On the other hand, UDP packet Flow Label values show more randomness. Because of its inherent variability, which would help to mask any anomalies brought about by encryption, this feature suggests that UDP may be a better option for embedding covert communications within the Flow Label field. 

\subsubsection{Special Case with ICMPv6}
ICMPv6 packets used for diagnostic purposes, such as Echo Requests and Replies, typically exhibit consistent or default Flow Label values, similar to TCP traffic. In contrast, ICMPv6 error messages often show implementation-dependent Flow Label values, which can sometimes appear random \cite{berger2020flaw}. This inconsistency opens the possibility of using the Flow Label field in such packets as a covert communication channel with a lower risk of detection.

\subsubsection{Encryption Strategy on Flow Label}
These features have led us to concentrate our covert communications efforts on error messages sent by UDP and ICMPv6. Using a Python script designed to create covert communication packets, we alter the Flow Label values in these packets. With the exception of the encrypted Flow Labels, the majority of these packets remain original to blend in with normal traffic. 
The Flow Label values are encoded using the RC4 encryption algorithm. The Flow Label's first eight bytes are used as a sequence identifier, and the final 16 bytes are encrypted. Both the RC4 encryption key and this sequence identifier are predetermined and decided upon by the parties involved in covert communication. A sequence arrangement might look like this:

\[
\text{sequence} = [E, A, 7, 1, 2, 3, 4, 5, 6, 8, 9, B, C, D, F, O]
\]

\noindent\textbf{Example of Encrypted Flow Labels:}
The list in Table \ref{tab:encrypted_Flow Labels} provides an example of how encrypted Flow Labels might appear in a stream of covert communications.

\begin{table}[h]
\captionsetup{width=0.8\linewidth, justification=centering}
\centering
\begin{tabular}{cc}
\toprule
\textbf{Packet} & \textbf{Encrypted Flow Label} \\
\midrule
1  & \texttt{EF1C0} \\
2  & \texttt{ACCDB} \\
3  & \texttt{785C1} \\
4  & \texttt{1D688} \\
5  & \texttt{2C488} \\
6  & \texttt{3C6C7} \\
7  & \texttt{4D3CD} \\
8  & \texttt{5D7DC} \\
9  & \texttt{685CB} \\
10 & \texttt{8CAC5} \\
11 & \texttt{9C8DD} \\
12 & \texttt{BCBC1} \\
13 & \texttt{CC6C9} \\
14 & \texttt{DD1C1} \\
15 & \texttt{FCAC6} \\
\bottomrule
\end{tabular}
\caption{Example of encrypted Flow Label values in covert communication packets.}
\label{tab:encrypted_Flow Labels}
\end{table}

A portion of the hidden message corresponds to each encrypted Flow Label. The sequence number and the strength of the RC4 encryption ensure that even if the packets are sent in a different order, the recipient will still be able to understand and interpret the message.  For instance, ``This is a covert communication'' is translated by the encrypted Flow Label values in Table \ref{tab:encrypted_Flow Labels}. 

\subsection{Length Field}

Since the payload length is specified by the IPv6 length field, which is not used in checksum calculations, it offers a chance to embed covert communications. This property makes it possible to make changes that are difficult to find using common integrity checks. 

\subsubsection{Encryption Strategy on Length Field}
We investigate whether the RC4 stream cipher can encrypt the Length field. To improve security, this method entails changing the data's ASCII values and using RC4 encryption. Adjusting ASCII values before encryption masks predictable patterns in the data, making unauthorized decryption more difficult. 

\subsubsection{Key Management}
Both parties engaged in covert communication must share a secret key in order for encryption and decryption to be effective. To guarantee the security of the communication, this key, which is essential for starting the encryption algorithm, needs to be closely guarded. 

\subsection{Address Space}
Our security analysis included investigating the possibility of using the IPv6 address space for covert communication. The last 8 bytes of an IPv6 address, which are the least important bits and usually contain the interface identifier or a portion, are where we specifically looked at whether information could be encoded. This technique guarantees that packets are routed correctly within the network by enabling address modifications without compromising the network prefix.

\subsubsection{Encryption Strategy on Address Space}
The RC4 algorithm is used to encrypt the least significant bits of the IPv6 address to preserve confidentiality. To prevent network routing disruptions, this encryption is restricted to the final 8 bytes of the address, leaving the remaining portion free to carry out its routing function. The communicating parties engaged in the covert exchange safely share the keys needed for this encryption. 

\subsubsection{Data Set Considerations}
Due to anonymization, only the first 8 bytes of the IPv6 addresses were accessible in our tests using the CAIDA 2019 anonymized dataset. Thus, in order to accomplish our study's objectives, the final 8 bytes had to be artificially generated. For packet sequences where the source and destination addresses' first eight bytes were consistent, indicating a connection between them, we made sure that the last eight bytes of these addresses, which were artificially generated, remained consistent as well, in order to imitate realistic network interactions and maintain consistency. This methodology proved instrumental in maintaining the integrity of our experiments and facilitating the investigation of surreptitious communication techniques in the IPv6 address space. 

\section{Experimental Study}

The experimental configuration used to evaluate the feasibility and detectability of covert communications in IPv6 networks is described in detail in this section. The purpose of our experiments was to examine different encryption strategies applied to different IPv6 fields by integrating covert packets into legitimate network traffic.

\subsection{Dataset Creation}

The datasets for our experiments were meticulously constructed to simulate real-world network traffic conditions. The Flow Label, Address Space, and Length fields were used to generate and carefully insert covert packets, which were encrypted based on the specified strategies.

\subsubsection{Covert Packet Integration}
\begin{itemize}
    \item \textbf{Flow Label and Length Field Encryption:} Covert packets were inserted into active, valid connections in which encryption was applied to either the Flow Label or the Length field. This method reduced the detectability of the covert packets among regular traffic by ensuring that they followed logical TCP sequence numbers and shared the same source and destination addresses. It is noteworthy that Flow Label encryption was applied to UDP and ICMPv6 error messages, but was specifically avoided in TCP packets due to their detectability. 
    
    \item \textbf{Address Space Encryption:} When the encryption strategy involved modifying the Address Space, the insertion of covert packets did not require alignment with existing traffic flows since the very nature of the encryption could mask the origin effectively. This method allowed more flexibility in packet insertion without the need for matching TCP sequence numbers or other header fields that might link packets to a specific flow.
\end{itemize}

To create covert communication packets and smoothly incorporate them into the dataset, We took inspiration from pcapStego and created a Python script to automate the process. The dataset underwent a systematic division, with 75\% of it being used for training and 25\% for testing. The split was carried out in an unrandomized manner to preserve the integrity of sequential data patterns.

\subsubsection{Covert Communication Challenges}
The primary challenge was ensuring that the covert packets remained undetectable. By using established connections to insert covert packets (except in cases of Address Space encryption), we aimed to mimic legitimate traffic patterns closely. This method tested the effectiveness of the encryption techniques and the ability of potential detection systems to identify anomalies within what appears to be regular traffic.

\subsection{Machine Learning Models}

Several machine learning models were used in our study to identify and categorize covert communications in IPv6 networks.  Both Covert Communication Type Identification (Multiclass Classification) and Covert Communication Detection (Binary Classification) were major tasks that these models played a key role in addressing. 

\begin{figure}[H]
\centering
\includegraphics[width=0.7\textwidth]{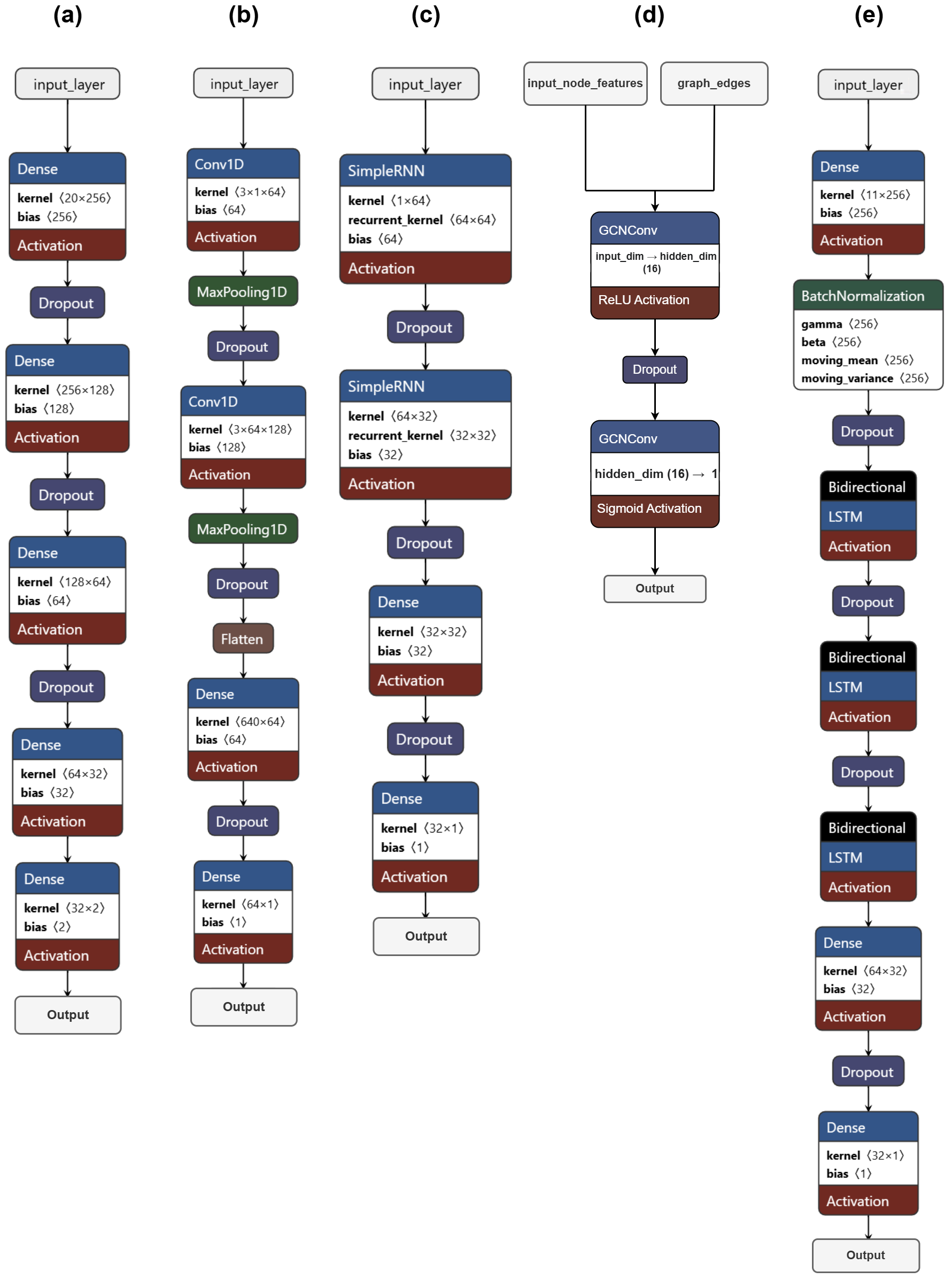}
\caption{Schematic of the neural network architectures explored: (a) a fully connected DNN, (b) a CNN with 1D convolutions, (c) an RNN based on SimpleRNN cells, a (d) GCN and (e) a bidirectional LSTM network.}
\label{fig:nn_architectures}
\end{figure}

\textbf{Decision Tree-Based Models:}
\begin{itemize}
    \item \textit{Random Forest}: Effective for binary and multiclass classification tasks, this model uses 100 decision trees and is set up with class weights to counteract imbalances between classes \cite{farnaaz2016random}. 
    \item \textit{Gradient Boosting and XGBoost}: By concentrating on fixing mistakes from earlier trees, both models seek to improve predictions progressively. XGBoost, specifically, has been optimized to minimize logarithmic loss across classes \cite{bhati2021improved,douiba2023improved}. 
    \item \textit{LightGBM}: It is especially well-suited for covert communication analysis because of its reputation for efficiency with large datasets and its application of class weights to enhance the handling of imbalanced data \cite{liu2021fast}.
\end{itemize}
\textbf{Neural Network Models:}
\begin{itemize}
    \item \textit{Deep Neural Network (DNN)}: Comprises four main layers: 1 dense layer of 256 units, followed by dropout, a second dense layer of 128 units, another dropout, and additional dense layers reducing to 2 units for output, designed to extract deep features from complex data patterns \cite{lansky2021deep}.
    \item \textit{Convolutional Neural Network (CNN)}: Includes two convolutional layers and pooling layers, followed by dropout layers and a flatten layer, culminating in dense layers for classification, particularly effective in pattern recognition tasks \cite{mohammadpour2022survey}.
    \item \textit{Recurrent Neural Network (RNN)} and \textit{Long Short-Term Memory (LSTM)}: The RNN uses two SimpleRNN layers, each followed by dropout, while the LSTM incorporates a bidirectional layer post several dense and dropout layers, enhancing the model’s ability to remember information over extended sequences \cite{sheikhan2012intrusion,imrana2021bidirectional}.
    \item \textit{Graph Convolutional Network (GCN)}: Applies convolutional processing to graph-structured data, ideal for network topology analysis \cite{deng2022flow}.
\end{itemize}

These models were rigorously tested to ensure their effectiveness under varied conditions, enhancing our detection systems' robustness and reliability against covert communications.

\subsection{Evaluation Metrics}

To rigorously evaluate the machine learning models' performance in detecting and classifying covert communications, we utilized several key metrics: Precision, Recall, $F_{1}$ Score, and Accuracy. These metrics provide insights into the models' effectiveness in terms of error minimization and information retrieval capabilities.

\subsubsection{Precision} 
Precision is the ratio of correctly predicted positive observations to the total predicted positives. It is a measure of a classifier's exactness. Higher precision relates to a lower false positive rate. The precision is mathematically defined as:
\begin{equation}
\text{Precision} = \frac{TP}{TP + FP}
\label{eq:precision}
\end{equation}
where \( TP \) is the number of true positives and \( FP \) is the number of false positives.

\subsubsection{Recall} Recall (also known as sensitivity) is the ratio of correctly predicted positive observations to all observations in actual classes. It is a measure of a classifier's completeness. The higher the recall, the more cases the classifier covers. The mathematical formula for the recall is:
\begin{equation}
\text{Recall} = \frac{TP}{TP + FN}
\label{eq:recall}
\end{equation}
where \( TP \) is the number of true positives and \( FN \) is the number of false negatives.

\subsubsection{$F_{1}$ Score} The $F_{1}$ Score is the weighted average of Precision and Recall. Therefore, this score takes both false positives and false negatives into account. It is especially useful when the class distribution is uneven. The formula for the $F_{1}$ score is:
\begin{equation}
\text{$F_1$ Score} = 2 \cdot \frac{\text{Precision} \cdot \text{Recall}}{\text{Precision} + \text{Recall}}
\label{eq:f1_score}
\end{equation}

\subsubsection{Accuracy} Accuracy is the ratio of correctly predicted observations to the total observations. It is useful when the target classes are well balanced. The formula for accuracy is:
\begin{equation}
\text{Accuracy} = \frac{TP + TN}{TP + TN + FP + FN}
\label{eq:accuracy}
\end{equation}
where \( TN \) is the number of true negatives.

Each of these metrics provides different insights into the performance of the models, helping to ensure that our evaluations are both comprehensive and informative.

\subsection{Generative AI Assisted Refinement}

\begin{figure*}[h!]
\centering
\includegraphics[width = 0.8\textwidth]{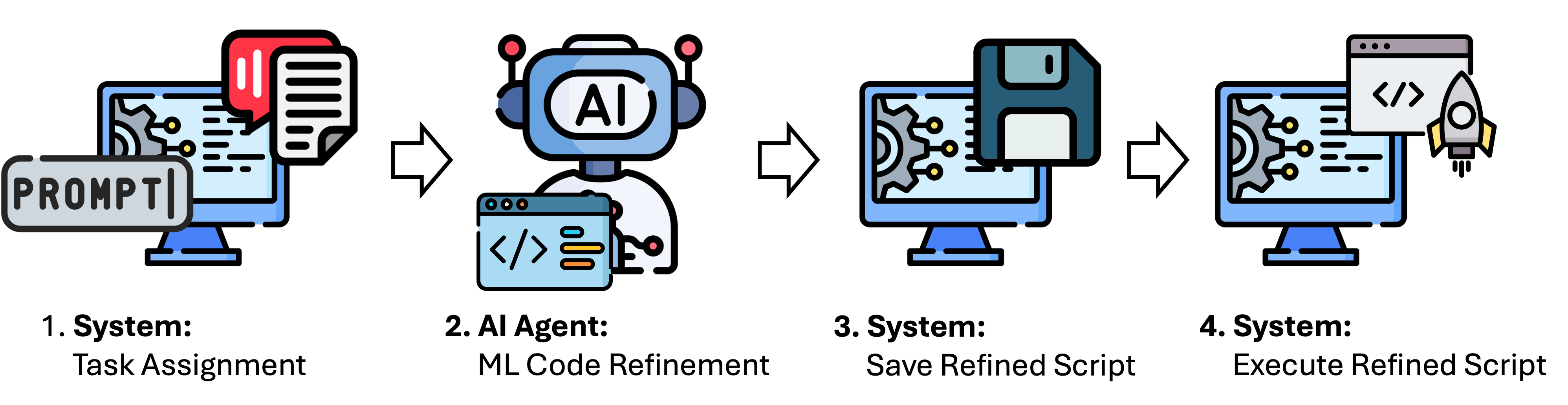}
\caption{AI Agent-Driven Script Refinement Pipeline}
\label{fig:genAI_Pipeline}
\end{figure*}

In this work, we refined Python scripts related to covert communication detection using the GPT-4-turbo model as a generative AI agent expert in code generation \mbox{\cite{chen2023teaching,yang2024swe}}. Utilizing generative AI, the classifier models underwent iterative refinement, with a binary classifier for detection and a multi-class classifier for identifying the type of covert communication. In accordance with the approach outlined by \mbox{\cite{rahman2025multi}}, the generative AI agents aimed to iteratively refine the Python scripts until their performance surpassed a specified threshold. Figure \mbox{\ref{fig:genAI_Pipeline}} illustrates the process.

The generative AI agent was configured with a maximum token limit of 4096, providing ample capacity to process and refine the scripts. An essential feature of this refinement process is that the AI agent can choose the most suitable machine learning or large language model (LLM) for classification tasks. This adaptability guarantees that the context and the data under examination are selected as the best strategy. Through user prompts, the agent is also given the accuracy threshold for classification, directing the refinement process toward meeting the necessary performance metrics.


The task assignment includes detailed information about the dataset, such as column properties, specific IPv6 fields vulnerable to covert communication, and other contextual information that can affect the detection process. The user prompt contains the path to the dataset, classification results (provided as data frames), Python scripts that need to be refined, and the classification accuracy threshold. 
Generative AI makes systematic improvements by iteratively reviewing, adjusting, and optimizing the classifiers until they reach or exceed the accuracy threshold. This process of refinement leads to improved detection and classification accuracy in cybersecurity applications by effectively automating model selection and script improvements using the capabilities of GPT-4-turbo.

The task assignment for the AI agent is shown as follows:
\begin{formal}
This work focuses on refining an existing Python-based detection pipeline (\texttt{covert\_ipv6.py}) for identifying and classifying covert communication within IPv6 traffic. The dataset, located at \texttt{dataset.csv}, includes labeled IPv6 packets with features extracted from various header fields, such as \texttt{Flow Label}, \texttt{AddressSpacePattern}, \texttt{PayloadLength}, and others. The classification results, including binary and multiclass outcomes, are stored in \texttt{results.txt}, while domain-specific knowledge about covert channel strategies is available in \texttt{knowledge.txt}.

The refinement process involves the following objectives:
\begin{enumerate}
    \item \textbf{Code Enhancement:} Improve the detection and classification logic in \texttt{covert\_ipv6.py}, particularly feature engineering, data preprocessing, and classifier selection.
    
    \item \textbf{Binary Classification:} Detect whether a given IPv6 packet is part of a covert channel or not. This involves improving the model's accuracy, precision, recall, and F1-score.
    
    \item \textbf{Multiclass Classification:} For packets classified as covert, further categorize them into specific covert types based on their manipulation strategies (e.g., Flow Label-based, LengthField-based).
    
    \item \textbf{Knowledge-Guided Feature Augmentation:} Incorporate human knowledge from \texttt{knowledge.txt} to derive new features, rules, or validation criteria. Examples include entropy of field values or irregular sequencing of header bytes.
    
    \item \textbf{Evaluation and Visualization:} Quantitatively evaluate the improvements via standard classification metrics and visualize the performance using confusion matrices and ROC curves.
\end{enumerate}

This multi-stage refinement pipeline aims to enhance both the robustness and explainability of covert IPv6 communication detection systems.
\end{formal}

\section{Results and Discussions}
This section presents a comprehensive evaluation of the machine learning models deployed to detect and classify covert communications in IPv6 traffic. The analysis is structured around two primary tasks: binary classification for identifying whether a packet contains covert communication and multiclass classification for determining the specific type of covert channel utilized. Standard performance metrics (including precision, recall, $F_{1}$ score, and precision) are used to assess the effectiveness of each model.

\begin{figure*}[h!]
\centering
\includegraphics[width=0.9\textwidth]{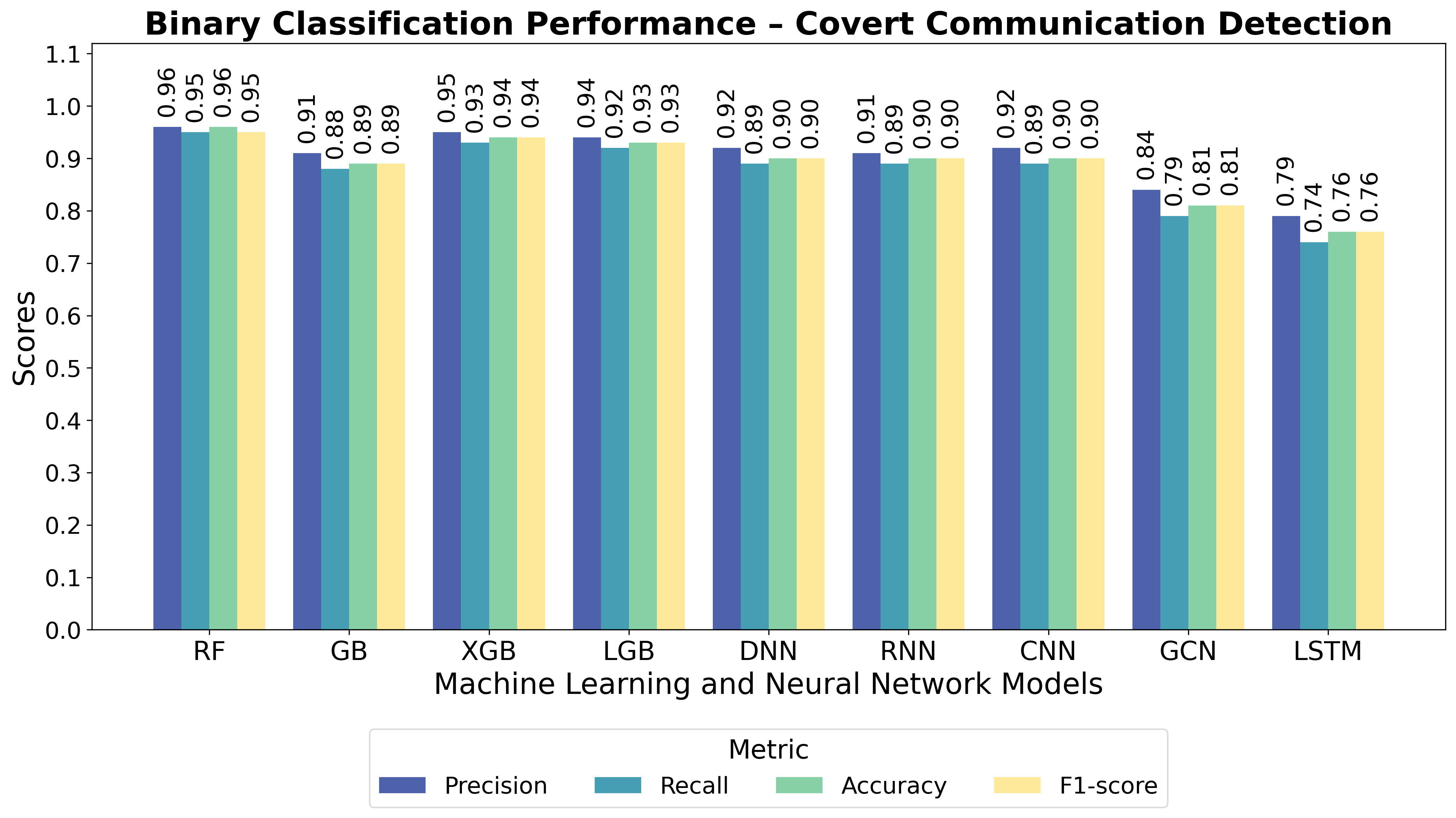}
\caption{Performance of Machine Learning Models on Covert Communication Type Identification Task}
\label{fig:binary_classification}
\end{figure*}

\begin{figure*}[h!]
\centering
\includegraphics[width=0.9\textwidth]{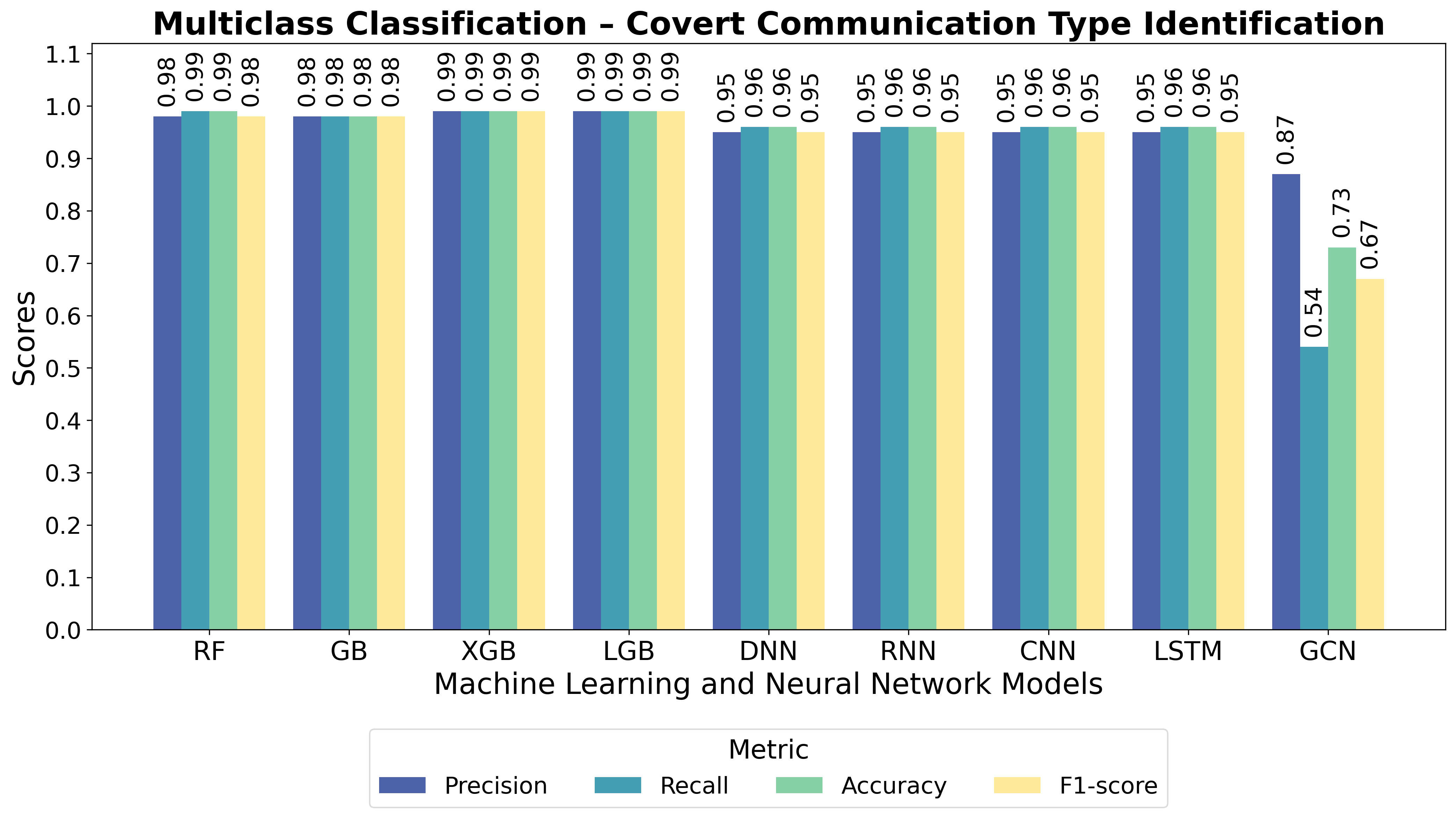}
\caption{Performance of Machine Learning models on Covert Communication Type Identification Task}
\label{fig:multiclass_classification}
\end{figure*}

Figures~\ref{fig:binary_classification} and~\ref{fig:multiclass_classification} illustrate comparative performance between models using bar graph visualizations. The analysis encompasses traditional ensemble methods including RF, GB, XGB, and LGB, as well as neural network architectures such as DNN, CNN, RNN, and GCN. The following subsections detail the performance of each model in the two classification tasks, providing insights into their relative strengths, limitations, and suitability for different detection scenarios.

\subsection{Binary Classification}

Among the evaluated models, Random Forest (RF), Extreme Gradient Boosting (XGB), and Light Gradient Boosting (LGB) showed consistent performance across multiple evaluation metrics. In particular, RF and XGB achieved relatively higher accuracy and $F_{1}$ scores compared to the other methods, indicating their suitability for the binary classification task in this context.

In contrast, the Graph Convolutional Network (GCN) exhibited lower performance across all evaluation metrics, suggesting limited effectiveness in the binary classification setting. This outcome may be attributed to the model’s dependence on non-Euclidean data structures, which are less prominent in the current task. Similarly, the Bidirectional Long Short-Term Memory (BiLSTM) model yielded lower scores relative to other neural network architectures, indicating that its performance in this context may be constrained and could benefit from further optimization.

\subsection{Multiclass Classification}

In the multiclass classification task, which presents a higher level of complexity, both the Random Forest (RF) and Extreme Gradient Boosting (XGB) models maintained relatively strong performance, suggesting their adaptability to multi-class scenarios. The Light Gradient Boosting (LGB) model achieved the highest accuracy and $F_{1}$ scores among all evaluated models, indicating its effectiveness in addressing the increased complexity of multiclass classification.

Compared the multiclass classification performance with the binary classification task, the Bidirectional LSTM (BiLSTM) demonstrated improved results in the multiclass classification setting. This relative improvement may be associated with the model’s ability to capture sequential dependencies, which can be beneficial in distinguishing between covert communication types that involve ordered patterns. For instance, in cases such as Hop Limit-based encoding, the BiLSTM appears to be effective in modeling the temporal variations in hop limit values, contributing to more accurate class differentiation. These findings suggest that sequence modeling architectures like BiLSTM may offer advantages in tasks where temporal or structural dependencies are relevant.

\begin{table}[ht]
\centering

\label{tab:combined_performance}
\begin{tabular}{@{}lcccc@{}}
\toprule
\textbf{Model} & \multicolumn{2}{c}{\textbf{Binary Classification}} & \multicolumn{2}{c}{\textbf{Multiclass Classification}} \\ 
\cmidrule(lr){2-3} \cmidrule(lr){4-5}
               & Accuracy & $F_{1}$ Score & Accuracy & $F_{1}$ Score \\ 
\midrule
RF & 0.96 & 0.95 & 0.99 & 0.98 \\
GB & 0.89 & 0.89 & 0.98 & 0.98 \\
XGB & 0.94 & 0.94 & 0.99 & 0.99 \\
LGB & 0.93 & 0.93 & 0.99 & 0.99 \\
DNN & 0.90 & 0.90 & 0.96 & 0.95 \\
RNN & 0.90 & 0.90 & 0.96 & 0.95 \\
CNN & 0.90 & 0.90 & 0.96 & 0.95 \\
GCN & 0.81 & 0.81 & 0.73 & 0.67 \\
LSTM & 0.76 & 0.76 & 0.96 & 0.95 \\
\bottomrule
\end{tabular}
\caption{Model Performance: Accuracy and $F_{1}$ Score for Binary and Multiclass Classification}
\label{tab:combined_performance}
\end{table}

However, the GCN showed a further decrease in performance in the multiclass classification setting, suggesting limited applicability in this context. This may be due to the model design, which is optimized for graph-structured data rather than the tabular feature-based representations used in covert communication detection tasks. In addition, DNN, CNN, and RNN exhibited stable and comparable performance, with CNN achieving slightly better results among neural network-based models. Although these architectures performed competitively, the improved performance of the Bi-LSTM highlights its relative strength in capturing sequential dependencies that are relevant to distinguish between different types of covert communication.

In general, traditional tree-based models, such as RF, XGB, and especially LGB, demonstrated exceptional performance in both binary and multiclass classification tasks, reaffirming their robustness for detection of covert communication. Neural network-based models also performed competitively, though GCN's architectural design constraints led to suboptimal results. Table~\ref{tab:combined_performance} summarizes the accuracy and $F_{1}$ scores for both tasks.

\section{Conclusion}

This study presents a systematic analysis of the potential use of various IPv6 header fields for covert communication. It also introduces machine learning-based approaches for binary detection and multiclass classification of covert traffic within IPv6 networks. Using a Python-based tool inspired by pcapstego, we curated a balanced data set aimed at machine learning applications that included normal and covertly modified IPv6 traffic. Our comparative analysis across various machine learning models, including RF, XGB, LGB, GB, DNN, CNN, RNN, GCN, and LSTM, reveals that Random Forest (RF) exhibited the most optimal performance, whereas the Graph Convolutional Network (GCN) showed the least favorable outcomes.

\subsection{Contributions and Distinguishing Features}
Our research addressed several critical challenges and presented novel solutions:
\begin{itemize}
    \item \textbf{Realistic Data Generation:} Unlike previous studies in this domain that often used unrealistic and overly simplistic datasets for covert communication, leading to near-perfect detection rates, our dataset generation approach introduces realistic complexities and variations. This approach better mimics actual operational environments, making the detection task more challenging and the results more indicative of real-world performance.
    \item \textbf{Comprehensive Machine Learning Evaluation:} We systematically evaluated a range of machine learning models, demonstrating that while traditional tree-based models like RF, XGB, and LGB perform robustly, the effectiveness of advanced neural network models like LSTM in handling multiclass tasks provides valuable insights into their deployment in network security.
\end{itemize}

\subsection{Future Directions}
Despite the promising results, the study highlights several areas for future research:
\begin{itemize}
    \item \textbf{Streaming Data Processing:} The current methods focus on accuracy rather than processing speed, which may limit real-world deployment for streaming data analysis. Future efforts should aim at optimizing these models for real-time data processing.
    \item \textbf{Adaptation to Diverse Network Environments:} The performance of machine learning models can degrade when deployed in different network settings from those they were trained on. Ongoing research should focus on developing adaptable models that can self-tune to varying network conditions without compromising accuracy.
    \item \textbf{Detection of Encrypted Covert Communications:} Further research is needed to enhance the detection capabilities for encrypted covert communications, which pose a significant challenge due to their sophisticated obfuscation techniques.
    \item \textbf{Integration of Generative AI: } Because generative AI generates and refines detection scripts dynamically based on changing threat landscapes, it can help with both the detection and classification phases. This robust tool can be used to counter covert communications because of its capacity to process and synthesize large amounts of packet data, identify subtle anomalies in IPv6 fields, and iteratively improve detection methods. Furthermore, the ability to select from a variety of machine learning models or extensive language models for classification allows the generative AI agent to swiftly adjust to newly discovered covert channel types. The integration of AI-driven real-time monitoring systems, in which generative AI creates and modifies detection algorithms on the fly based on real-time network traffic, may be the subject of future research. This would improve the overall resilience of cybersecurity defenses by enabling dynamic adaptation to new attack vectors in IPv6 environments. To further enhance the process of detecting and classifying covert communications, multi-agent systems that leverage the collaboration of various generative AI models to exchange insights on vulnerabilities may be employed.

\end{itemize}

In conclusion, this work improves the detection and classification of covert communications within IPv6 networks by combining protocol field analysis with machine learning techniques. The data set and methodology used aim to reflect realistic scenarios, supporting future research on practical and adaptable cybersecurity approaches.


\section*{Declarations}
This study makes use of the \textit{2019 CAIDA IPv6 Launch Day Anonymized Internet Traces} for normal network traffic. Access to these traces is subject to CAIDA’s usage agreement and is not publicly available; interested researchers may request access via the CAIDA website \href{https://www.caida.org}{(https://www.caida.org)}.

To support the reproducibility of our experiments, we have made representative samples publicly available in this \href{https://github.com/Waliboii/ipv6_covert/blob/main/sample_nyc_filtered_pcap_caida.csv}{GitHub repository}, including both normal traffic and covert traffic samples generated during the study.

\section*{Acknowledgment}
This work is supported by the National Science Foundation (NSF) under project numbers 1624668, 1921485, and 2335046; the U.S. Department of Energy–National Nuclear Security Administration (DOE-NNSA) under Award Number DE-NA0003946; and the AGILITY project (Grant No. 4263090), sponsored by the Korea Institute for Advancement of Technology (KIAT), South Korea. The authors would like to thank the Department of Systems and Industrial Engineering at the University of Arizona for matching collaboration opportunities with the National Security Agency through the Information Security Research and Education (INSuRE) research collaborative.

\ifCLASSOPTIONcaptionsoff
  \newpage
\fi

\bibliographystyle{IEEEtran} 
\bibliography{ref}

\end{document}